\documentclass{mpla}
\usepackage[super]{cite}
\usepackage{graphicx}

\def\etc{{\it etc.}}
\def\ie{{\it i.e.}}
\def\eg{{\it e.g.}}

\def\apriori{{\it a priori}}

\def\msolar{M_\odot}
\def\fstatistic{$\mathcal{F}$-statistic}

\begin{document}

\markboth{K. Riles}
{Recent Searches for Continuous Gravitational Waves}

%%%%%%%%%%%%%%%%%%%%% Publisher's Area please ignore %%%%%%%%%%%%%%
\catchline{}{}{}{}{}
%%%%%%%%%%%%%%%%%%%%%%%%%%%%%%%%%%%%%%%%%%%%%%%%%%%%%%%%%%%%%%%%%%%

\title{RECENT SEARCHES FOR \\ CONTINUOUS GRAVITATIONAL WAVES}

\author{\footnotesize KEITH RILES}

\address{Physics Department, University of Michigan, 450 Church Street\\
Ann Arbor, Michigan 48109-1040, USA\\
kriles@umich.edu}

\maketitle

\pub{Received (17 November 2017)}{Revised (Day Month Year)}

\begin{abstract}

  Gravitational wave astronomy opened dramatically in September 2015
  with the LIGO discovery of a distant and massive binary black hole coalescence.
  The more recent discovery of a binary neutron star merger, followed by
  a gamma ray burst and a kilonova, reinforces the excitement of this
  new era, in which we may soon see other sources of gravitational waves,
  including continuous, nearly monochromatic signals. Potential
  continuous wave (CW) sources include rapidly spinning galactic neutron 
  stars and more exotic possibilities, such as emission from axion Bose Einstein ``clouds''
  surrounding black holes. Recent searches in Advanced LIGO data are presented,
  and prospects for more sensitive future searches discussed.

\end{abstract}

%Are PACS still used?
%\ccode{PACS Nos.: include PACS Nos.}

\section{Introduction}	

LIGO's detection in September 2015 of gravitational waves from the coalescence 
of heavy stellar-mass black holes (GW150914)~\cite{bib:GW150914} inaugurated
observational gravitational wave astronomy, a new field ripe for discovery and
one already deepening our understanding of astrophysics. The subsequent detections of
additional binary black hole mergers
GW151226,~\cite{bib:GW151226} GW170104,\cite{bib:GW170104}, GW170608,~\cite{bib:GW170608} and GW170814,~\cite{bib:GW170814}
along with a binary neutron star merger~\cite{bib:GW170817} (last two events seen also by Virgo) associated with a short gamma ray burst (GRB)~\cite{bib:GW170817GRB}
and a kilonova~\cite{bib:GW170817MMA} have confirmed 
the excitement of detecting gravitational wave transients.

An entirely different type of gravitational wave
source merits attention too -- continuous waves (CWs) emitted by compact spinning objects, most
notably non-axisymmetric neutron stars in our own galaxy. Despite their relative closeness
($\sim$several kpc vs tens to hundreds of Mpc), such sources are expected to produce gravitational-wave
strain amplitudes orders of magnitude weaker than those seen from compact binary mergers, \ie, O($10^{-25}$) or smaller
compared to O($10^{-21}$). Our only hope of detecting such weak signals comes from integrating
data over long time spans, but such integrations incur, in most searches, substantial computing
cost from covering finely a large parameter space volume (\eg, frequency evolution, sky location, potential orbital
parameters). In nearly every CW search, achievable sensitivity is limited by finite computational resources.

In the following, section~\ref{sec:sources} discusses both conventional and exotic potential sources
of CW gravitational radiation, and section~\ref{sec:searches} presents a wide variety of searches tailored
to {\it a priori} knowledge (or lack thereof) of the source properties, along with results (so far all negative)
of those searches. Where available, results from recent searches in Advanced LIGO data will be presented; for ongoing
searches not yet published, older results from Initial LIGO and Virgo searches will be presented instead.
Finally, section~\ref{sec:future} discusses the prospects for future searches in Advanced LIGO and Virgo data,
as detector sensitivities and algorithms improve.

This brief review focuses on CW radiation potentially detectable with current-generation
ground-based gravitational wave interferometers, which are sensitive in the audio band.
Past and future searches for lower-frequency CW radiation from supermassive black hole binaries
at $\sim$nHz frequencies using pulsar timing arrays~\cite{bib:ptareview} or from stellar-mass galactic binaries
at $\sim$mHz frequencies using the space-based LISA~\cite{bib:LISA} are not discussed here. 

\section{Potential CW Sources}
\label{sec:sources}

In the frequency band of present 
ground-based detectors, the canonical sources of continuous gravitational waves are
galactic, non-axisymmetric neutron stars spinning fast enough that
their rotation frequencies (or twice their frequencies) are in the LIGO and Virgo detectable band. 
These nearby neutrons stars
offer a ``conventional'' source of CW radiation -- as astrophysically extreme as
such objects are.

A truly exotic postulated source
is a ``cloud'' of bosons, such as QCD axions, surrounding a fast-spinning
black hole, bosons that can condense in gargantuan numbers to a small
number of energy levels, enabling coherent
gravitational wave emission from boson annihilation or from level transitions.
Attention here focuses mainly on the conventional neutron stars, but
the exotic boson cloud scenario is also discussed briefly. 

\subsection{Fast-spinning neutron stars}
\label{sec:neutronstarsources}

Spinning neutron stars could generate detectable gravitational waves
via several possible mechanisms.
Isolated neutron stars may exhibit intrinsic non-axisymmetry from residual crustal
deformation (\eg, from cooling \&\ cracking of the crust),~\cite{bib:crustdeformation} from
non-axisymmetric distribution of magnetic field energy trapped
beneath the crust~\cite{bib:buriedbfieldasymmetry} or from a pinned neutron superfluid component
in the star's interior (see Ref.~\refcite{bib:haskelletal} for a discussion of emission from
magnetic and thermal ``mountains'' and Refs.~\refcite{bib:lasky,bib:GandG} for recent, comprehensive reviews of
GW emission mechanisms from neutron stars). Maximum allowed asymmetries depend on the
neutron star equation of state~\cite{bib:johnsonmcdanielowen} and on the breaking strain of
the crust~\cite{bib:horowitz}.

Normal modes of oscillations could also arise, including
$r$-modes in which quadrupole mass currents emit gravitational waves~\cite{bib:rmodes1,bib:rmodes2,bib:rmodes3,bib:rmodes4}.
These $r$-modes can be inherently unstable, arising from azimuthal
interior currents that are retrograde in the star's rotating frame, but
are prograde in an external reference frame. As a result, the
quadrupolar gravitational wave emission due to these currents
leads to an {\it increase} in the strength of the current. This
positive-feedback loop leads to a potential intrinsic instability.
The frequency of such emission is expected to be approximately 4/3 the rotation 
frequency~\cite{bib:rmodes1,bib:rmodes2,bib:rmodes3,bib:rmodes4}.  Serious
concerns have been raised,~\cite{bib:rmodesdoubts,bib:GandG} however, about the 
detectability of the emitted radiation for young 
isolated neutron,
where mode saturation appears 
to occur at low $r$-mode amplitudes because
of various dissipative effects. A recent study,~\cite{bib:alfordschwenzeryoungpulsar} however,
is more optimistic about newborn neutron stars. The same authors, on the other hand, find that $r$-mode emission
from millisecond pulsars is likely to be undetectable by Advanced LIGO~\cite{bib:alfordschwenzerMSP}.
The notion of a runaway rotational instability was
first appreciated for high-frequency $f$-modes,~\cite{bib:cfs1,bib:cfs2} (Chandrasekhar-Friedman-Schutz
instability), but realistic viscosity effects seem likely to 
suppress the effect in conventional neutron star production~\cite{bib:cfskiller1,bib:cfskiller2}.
The $f$-mode stability could play an important role, however, for
a supramassive or hypermassive neutron star formed as the remnant
of a binary neutron star merger~\cite{bib:donevaetal}.

In addition, as discussed below, a binary neutron star may experience direct non-axisymmetry
from non-isotropic accretion~\cite{bib:owenelastic,bib:ushomirskyetal,bib:melatospayne} (also possible for an isolated 
young neutron star that has experienced fallback accretion shortly
after birth), or may exhibit $r$-modes induced by accretion spin-up.

Given the various potential mechanisms for generating continuous gravitational waves
from a spinning neutron star, detection of the waves should yield valuable information
on neutron star structure and on the equation of state of nuclear matter 
at extreme pressures, especially when combined with electromagnetic observations
of the same star.

In principle, there should be O(10$^{8-9}$) neutron stars in our galaxy,~\cite{bib:nspopulation}
out of which only about 2500 have been detected, primarily as radio pulsars.
This small fraction of detections is expected, for several reasons.
Radio pulsations require high magnetic field and rotation frequency. An early study~\cite{bib:deathline,bib:pulsarastronomy}
implied the relation $B\cdot f_{\rm rot}^2>1.7\times10^{11}$ G$\cdot$(Hz)$^2$ based on a model
of radiation dominated by electron-positron pair creation in the stellar magnetosphere, a model
broadly consistent with empirical observation, although the resulting ``death line''
in the plane of period and period derivative is perhaps better understood
to be a valley.~\cite{bib:ZhangHardingMuslimov}.
%Radio pulsations appear empirically to require the combination of the magnetic field and
%rotation frequency to satisfy the approximate 
%relation $B\cdot f_{\rm rot}^2>1.7\times10^{11}$ G$\cdot$(Hz)$^2$~\cite{bib:deathline}.
As a result, isolated pulsars seem to have pulsation lifetimes of O($10^7$ years),~\cite{bib:pulsarastronomy}
after which they are effectively radio-invisible.
On this timescale, they also cool to where thermal X-ray emission is difficult
to detect. There remains the possibility of X-ray emission
from steady accretion of interstellar medium (ISM), but it appears that the
kick velocities from birth highly suppress such accretion~\cite{bib:bondi1,bib:bondi2}
which depends on the inverse cube of the star's velocity through
the ISM.

A separate population of pulsars and non-pulsating neutron stars
can be found in binary systems. In these binaries accretion from a 
non-compact companion star can lead to ``recycling,'' in which
a spun-down neutron star regains angular momentum from the infalling matter.
The rotation frequencies achievable through this spin-up are
impressive -- the fastest known rotator is J1748-2446ad at 716 Hz~\cite{bib:hessels}.
One class of such systems is the set of low mass X-ray binaries (LMXBs)
in which the neutron star ($\sim$1.4 $\msolar$) has a much
lighter companion ($\sim$0.3 $\msolar$)~\cite{bib:pulsarastronomy} that
overfills its Roche lobe, spilling material onto an accretion
disk surrounding the neutron star or possibly spilling material
directly onto the star, near its magnetic polar caps. 
When the donor companion star eventually shrinks and decouples
from the neutron star, the neutron star can retain a large
fraction of its maximum angular momentum and rotational energy.
Because the neutron star's magnetic field decreases during
accretion (through processes that are not well understood),
the spin-down rate after decoupling can be very small.

Equating the rotational energy loss rate to magnetic dipole
radiation losses, leads to the relation~\cite{bib:pacini}:
\begin{equation}
\left({dE\over dt}\right)_{\rm mag} \quad = \quad -{\mu_0M_\perp^2\omega^4\over6\pi c^3},
\end{equation}
where $\omega$ is the rotational angular speed,
$M_\perp$ is the component of the star's magnetic dipole
moment perpendicular to the rotation axis: $M_\perp=M\sin(\alpha)$,
with $\alpha$ the angle between the axis and north magnetic pole.
In a pure dipole moment model, the magnetic pole field strength
at the surface is $B_0 = \mu_0M\,/\,2\pi R^3$.
Equating this energy loss to that of the (Newtonian) rotational
energy ${1\over2}I_{\rm zz}\omega^2$ leads to the prediction:
\begin{equation}
{d\omega\over dt} \quad = \quad -{\mu_0R^6\over6\,\pi c^3I_{\rm zz}}B_\perp^2\omega^3.
\end{equation}
Note that the spin-down rate is proportional to the square of $B_\perp=B_0\sin(\alpha)$
and to the cube of the rotation frequency.

More generally, the spin-down rate is often taken to have a power law dependence: $\dot f = Kf^n$
for some negative constant $K$ and exponent $n$ that depends on the energy loss mechanism
(\eg, magnetic dipole emission: $n=3$; gravitational radiation: $n=5$ or $n=7$, depending on mechanism). In this approximation,
the age $\tau$ of a neutron star can be related to its birth rotation frequency $f_0$ and
current frequency $f$ by ($n\ne1$):
\begin{equation}
\tau \quad = \quad -\left[{f\over (n-1)\,\dot f}\right]\,\left[1-\left({f\over f_0}\right)^{(n-1)}\right],
\end{equation}
and in the case that $f\ll f_0$, 
\begin{equation}
\label{eqn:approxageindex}
\tau \quad \approx \quad -\left[{f\over (n-1)\,\dot f}\right]\,.
\end{equation}
Assuming $n$ (often called the ``braking index'' and derived 
from the ratio $f\ddot f/\dot f^2$ when $\ddot f$ is measurable) is three
(magnetic dipole dominance), leads
to approximate inferred ages for many binary radio pulsars 
in excess of $10^9$ years and even well over $10^{10}$ years~\cite{bib:atnfdb}.
One calculation suggests that 
this surprising result can be explained by
reverse-torque spin-down during the Roche lobe
decoupling phase~\cite{bib:taurisrldp}. In fact,
measured braking indices for even young pulsars tend to be less than three,
suggesting that the model of a neutron star spinning down with
constant magnetic field is, most often, inaccurate~\cite{bib:pulsarastronomy}.
One recently reported exception, however, is X-ray pulsar J1640-4631
with a measured index of 3.15$\pm$0.03~\cite{bib:archibaldetal}.
See Ref.~\refcite{bib:palomba1,bib:palomba2} for discussions of spin-down evolution
in the presence of both gravitational wave and electromagnetic
torques. Other suggested mechanisms for less-than-3 braking indices are
re-emerging buried magnetic fields and
a decaying inclination angle between the magnetic dipole axis and
the spin axis~\cite{bib:tauriskonar,bib:emergingBfield,bib:decayinginclination}.
An interesting observation of the aftermath of two
short GRBs noted indirectly inferred braking indices near or equal to
three,~\cite{bib:magnetarbrakingindex}
suggesting the rapid spin-down of millisecond magnetars, possibly born
from neutron star mergers. (No direct evidence of a such a post-merger remnant
has been observed from GW170817~\cite{bib:GW170817remnant}.)
It has been argued recently that the inter-glitch evolution of spin for the
X-ray pulsar J0537-6910 displays behavior consistent with a braking index of 7,~\cite{rmodeJ0537-6910}
which is expected for a spin-down dominated by $r$-mode emission.

Speaking broadly, one can identify three categories
of neutron stars that are potentially detectable via continuous
gravitational waves: relatively young, isolated stars with spin
frequencies below $\sim$50 Hz, such as the Crab pulsar; 
actively accreting stars in binary systems; and recycled
``millisecond'' stars for which accretion has ceased and which generally
have rotation frequencies above 100 Hz. In some cases
the companion donor has disappeared, \eg, via ablation, 
leaving an isolated neutron star, but most known millisecond
pulsars remain in binary systems~\cite{bib:atnfdb}.

Let's now consider the gravitational radiation one might
expect from these stars. If a star at a distance $r$ away has a quadrupole
asymmetry, parametrized by its ellipticity:
\begin{equation}
\label{eqn:ellipticity}
\epsilon \quad \equiv \quad {I_{xx}-I_{yy}\over I_{zz}}, 
\end{equation}
and if the star is spinning about the approximate symmetry axis of rotation ($z$),
(assumed optimal -- pointing toward the Earth), then the expected intrinsic strain amplitude $h_0$ is
\begin{equation}
\label{eqn:hexpected}
h_0 \quad = \quad {4\,\pi^2GI_{\rm zz}f_{\rm GW}^2\over c^4r}\,\epsilon 
\quad = \quad (1.1\times10^{-24})\left({I_{\rm zz}\over I_0}\right)\left({f_{\rm GW}\over1\>{\rm kHz}}\right)^2
\left({1\>{\rm kpc}\over r}\right)\left({\epsilon\over10^{-6}}\right),
\end{equation}
where $I_0=10^{38}$ kg$\cdot$m$^2$ (10$^{45}$ g$\cdot$cm$^2$) is a nominal moment of inertia of
a neutron star, and the gravitational radiation is emitted at frequency $f_{\rm GW}=2\,f_{\rm rot}$.
The total power emission in gravitational waves from
the star (integrated over all angles) is 
\begin{equation}
\label{eqn:powerloss}
{dE\over dt} \quad = \quad - {32\over5} {G\over c^5}\,I_{\rm zz}^2\, \epsilon^2\, \omega^6 
\quad = \quad- (1.7\times10^{33}\>{\rm J/s})\left({I_{\rm zz}\over I_0}\right)^2
\left({\epsilon\over10^{-6}}\right)^2
\left({f_{\rm GW}\over1\>{\rm kHz}}\right)^6.
\end{equation}
For an observed neutron star of measured $f$ and $\dot f$,
one can define the ``spin-down limit'' on maximum detectable
strain by equating the power loss in equation~(\ref{eqn:powerloss})
to the time derivative of the (Newtonian) rotational kinetic
energy: ${1\over2}I\omega^2$, as above for magnetic dipole radiation. 
One finds:
\begin{eqnarray}
\label{eqn:spindownlimit}
h_{\rm spin-down}\quad & = & \quad {1\over r}\sqrt{-{5\over2}{G\over c^3}I_{\rm zz}{\dot f_{\rm GW}\over f_{\rm GW}}} \nonumber \\
& = & \quad (2.5\times10^{-25})\left({1\>{\rm kpc}\over r}\right)\sqrt{\left({1\>{\rm kHz}\over f_{\rm GW}}\right)
\left({-\dot f_{\rm GW}
\over10^{-10}\>{\rm Hz/s}}\right)\left({I_{\rm zz}\over I_0}\right)}.
\end{eqnarray}
Hence for each observed pulsar with a measured frequency spin-down and
distance $r$,
one can determine whether or not energy conservation even permits detection
of gravitational waves in an optimistic scenario. Unfortunately,
nearly all known pulsars have strain spin-down limits below what
can be detected by the LIGO and Virgo detectors at current
sensitivities, as discussed below. 

A similarly optimistic limit based only on the age of a known neutron 
star of unknown spin frequency can also be derived. If one assumes a star is spinning down entirely
due to quadrupolar gravitational radiation, then the energy loss for this {\it gravitar} satisfies equation~(\ref{eqn:approxageindex})
with a braking index of five (seven for $r$-mode emission).
Assuming a high initial spin frequency,
the star's age then satisfies: $\tau_{\rm gravitar}\quad = \quad -{f\over 4\,\dot f}$.

If one knows the distance and the age of the star, \eg, from
the expansion rate of its visible nebula, then
under the assumption that the star has been
losing rotational energy since birth primarily
due to gravitational wave emission, then one
can derive the following frequency-independent
age-based limit on strain~\cite{bib:cwcasamethod}:
\begin{equation}
h_{\rm age} \quad = \quad (2.2\times10^{-24})\left({1\>{\rm kpc}\over r}\right)\sqrt{\left({1000\>yr\over\tau}\right)
\left({I_{\rm zz}\over I_0}\right)}.
\end{equation}
A notable example is the Compact Central Object (CCO)
in the Cassiopeia A supernova remnant. Its birth
aftermath may have been observed by Flamsteed~\cite{bib:casabirth} in
1680, and the expansion of the visible shell is consistent
with that date. Hence Cas~A, which is visible in X-rays 
but shows no pulsations, is almost certainly a very young 
neutron star at a distance of about 3.4 kpc. From the above equation,
one finds an age-based strain limit of $1.2\times10^{-24}$, which is accessible to
LIGO and Virgo detectors in their most sensitive band.

Derivations for similar indirect and age-based limits on the dimensionless $r$-mode emission amplitude $\alpha$
can be found in Ref.~\refcite{bib:owenalpha}.

A simple steady-state argument by Blandford~\cite{bib:thorne300} led to 
an early estimate of the maximum detectable strain amplitude expected from a population of
isolated gravitars of a few times 10$^{-24}$, independent of typical ellipticity values,
in the optimistic scenario that most neutron stars become gravitars. 
A later detailed numerical simulation~\cite{bib:knispelallen}
revealed, however, that the steady-state assumption does not generally hold, leading
to ellipticity-dependent expected maximum amplitudes that can be 2-3 orders
of magnitude lower in the LIGO band for ellipticities as low as 10$^{-9}$ and a few
times lower for ellipticity of about $10^{-6}$.

Another approximate empirically determined strain limit can be defined
for accreting neutron stars in binary systems,
such as Scorpius X-1. The X-ray luminosity  from
the accretion is a measure of mass accumulation rate at
the surface. As the mass rains down on the surface
it can add angular momentum to the star, which in 
equilibrium may be radiated away in gravitational waves.
Hence one can derive a torque-balance limit~\cite{bib:wagoner,bib:papapringle,bib:rmodes2}:
\begin{equation}
\label{eqn:torquebalance}
h_{\rm torque}\quad = \quad(5\times10^{-27})
\sqrt{\left({600\>{\rm Hz}\over f_{\rm GW}}\right)
\left({\mathcal{F}_{\rm x}\over10^{-8}\>{\rm erg/cm}^2/{\rm s}}\right)},
\end{equation}
where $\mathcal{F}_{\rm x}$ is the observed energy flux at the Earth of
X-rays from accretion. Note that this limit is independent
of the distance to the star.

The notion of gravitational wave torque equilibrium is potentially important,
given that the maximum observed rotation frequency of neutron
stars in LMXBs is substantially lower than one might expect from
calculations of neutron star breakup rotation speeds ($\sim$1400 Hz)~\cite{bib:breakupspeed}.
It has been suggested~\cite{bib:speedlimit} that there is a ``speed limit''
governed by gravitational wave emission that governs the maximum
rotation rate of an accreting star. In principle, the distribution
of frequencies could have a quite sharp upper frequency cutoff,
since the angular momentum emission is proportional to the 
5th power of the frequency. For example, for 
an equilibrium frequency corresponding to a particular accretion rate,
doubling the accretion rate would increase the equilibrium frequency
by only about 15\%. Note, however, that a non-GW speed limit may arise
from interaction between the neutron star's magnetosphere and an accretion disk~\cite{bib:ghoshlamb,bib:haskellpatruno,bib:patrunohaskelldangelo}.

A number of mechanisms have been proposed by which the accretion
leads to gravitational wave emission in binary systems. The simplest is localized accumulation
of matter, \eg, at the magnetic poles (assumed offset from the rotation axis), 
leading to a non-axisymmetry.
One must remember, however, that matter can and will diffuse into
the crust under the star's enormous gravitational field. This diffusion of
charged matter can be slowed by the also-enormous magnetic fields in
the crust, but detailed calculations~\cite{bib:vigeliusmelatos} indicate the
slowing is not dramatic. Another proposed mechanism is excitation of
$r$-modes in the fluid interior of the star,~\cite{bib:rmodes1,bib:rmodes2,bib:rmodes3,bib:rmodes4}
with both steady-state emission and cyclic spin-up/spin-down 
possible~\cite{bib:rmodeslmxb,bib:rmodesdoubts}. Intriguing,
sharp lines consistent with expected $r$-mode frequencies were reported recently
in the accreting millisecond X-ray pulsar XTE J1751$-$305~\cite{bib:strohmayermahmoodifar1}
and in a thermonuclear burst of neutron star 4U 1636--536~\cite{bib:strohmayermahmoodifar2}.
The inconsistency of the observed stellar spin-downs for these sources with ordinary r-mode
emission, however, suggests that a different type of oscillation is being observed~\cite{bib:anderssonjonesho}
or that the putative r-modes are restricted to the neutron star crust and hence gravitationally
much weaker than core r-modes~\cite{bib:lee2014}. Another recent study~\cite{bib:patrunohaskellandersson}
suggests that spin frequencies observed in accreting LMXB's are consistent with two sub-populations, where
the narrow higher-frequency component ($\sim$575 Hz with standard deviation of $\sim$30 Hz) may signal an equilibrium
driven by gravitational wave emission. 

\subsection{Axion clouds bound to black holes}

A provocative idea receiving increased theoretical attention in
recent years posits that dark matter is not only composed
of electromagnetically invisible massive bosons, such as axions, but
that such bosons could be disproportionately found in the vicinity of rapidly spinning black
holes~\cite{bib:axiversearvanitaki1,bib:axiversearvanitaki2}.
These bosons could, in principle, be spontaneously created via energy extraction from the
black hole's rotation~\cite{bib:penrose1,bib:penrose2} and form a Bose-Einstein ``cloud'' with
nearly all of the quanta occupying a relatively small number of energy levels. For a cloud bound to a black hole,
the approximate inverse-square law attraction outside the Schwarzchild radius leads to an
energy level spacing directly analogous to that of the hydrogen atom. 
The number of quanta occupying the low-lying levels can be amplified enormously by the phenomenon of superradiance
in the vicinity of a rapidly spinning black hole (angular momentum comparable to
maximum allowed in General Relativity). The bosons in a non-$s$ ($\ell>0$) negative-energy
state can be thought of as propagating
in a well formed between an $\ell$-dependent centrifugal barrier at $r>r_{\rm Schwarzchild}$ 
and a potential rising toward zero as $r\rightarrow\infty$,
where wave function penetration into the black hole ergosphere permits transfer of energy
from the black hole spin~\cite{bib:superradiance1,bib:superradiance2,bib:superradiance3} into the creation of new quanta.

Two particular gravitational wave emission modes of interest here can arise in the axion scenario, both
potentially leading to intense coherent radiation~\cite{bib:axionarvanitaki}.
In one mode, axions can annihilate with each other to produce gravitons with frequency double that
corresponding to the axion mass: $f_{\rm graviton} = 2m_{\rm axion}c^2/h$. In another mode,
emission occurs from level transitions of quanta in the cloud.
This Bose condensation is most pronounced when
the reduced Compton wavelength of the axion is comparable to but larger than the scale of the black hole's Schwarzchild radius:
\begin{equation}
 {\lambda\over2\,\pi} = {\hbar\over m_{\rm axion}c} \gtrapprox {2\,GM_{\rm BH}\over c^2} \qquad
  \Rightarrow \qquad m_{\rm axion} \lessapprox (7\times10^{-11} {\>{\rm ev/c^2}}) {\msolar\over M_{\rm BH}} ,
\end{equation}
\noindent where $\hbar$ and $G$ are Planck's and Newton's constants.
Given the many orders of magnitude
uncertainty in, for example, axion masses that could account for dark matter,\cite{bib:darkmatterreview}
the relatively narrow mass window accessible to currently feasible CW searches (1-2 orders of magnitude) makes
searching for such an emission a classic example of ``lamppost'' physics, where one can only
hope that nature places the axion in this lighted area of a vast parameter space. 

In principle, searching for these potential CW sources requires
no fundamental change in the search methods described below, but search optimization can be
refined for the potentially very slow frequency evolution expected during both annihilation and
level transition emission. In addition, for a known black hole
location, a directed search can achieve better sensitivity than an all-sky search.
Note that for string axiverse models, however,
the axion cloud~\cite{bib:axiversearvanitaki1,bib:axiversearvanitaki2,bib:axionjapanesegroup} can experience significant self-interactions which can
lead to appreciable frequency evolution of the signal and to uncertainty on that evolution,
a complication less important for the postulated QCD axion~\cite{bib:axionarvanitaki}.
One interesting suggestion includes the possibility that a black hole formed from the detected merger of binary black
holes could provide a natural target for follow-up CW searches~\cite{bib:axionBBHmerger}.
To date, no published searches have been tailored for a black hole axion cloud source,
but existing (non-optimized) limits on neutron star emission can be reinterpreted as limits on
such emission~\cite{bib:axionarvanitaki}. Recent studies~\cite{bib:brito1,bib:brito2} argue that 
the lack of detection of a stochastic gravitational radiation background from the superposition of
extragalactic black holes already places significant limits on axion masses relevant to CW searches.
  
\section{Search Methods and Results}
\label{sec:searches}

As discussed in section~\ref{sec:sources}, continuous-wave (CW) gravitational 
radiation detectable by ground-based detectors is expected most plausibly
from rapidly spinning neutron stars in our galaxy. Search strategies
for CW radiation depend critically upon the \apriori\ knowledge one
has about the source. It is helpful to classify CW searches into three
broad categories~\cite{bib:prixreview,bib:rilesppnp}: 1) {\it targeted} searches in which
the star's position and rotation frequency are known, \ie, known 
radio, X-ray or $\gamma$-ray pulsars; 
2) {\it directed} searches in which the star's position is known, but rotation
frequency is unknown, \eg, a non-pulsating X-ray source at the
center of a supernova remnant; and 3) {\it all-sky} searches for unknown
neutron stars. The volume of parameter space over which to search increases
in large steps as one progresses through these categories. In each
category a star can be isolated or binary. For 2) and 3) any unknown binary
orbital parameters further increase the search volume.
In all cases we expect (and have now verified from unsuccessful searches to date)
that source strengths are very small. Hence one must integrate data over long
observation times to have any chance of signal detection. How much one knows about
the source governs the nature of that integration. In general, the greater that
knowledge, the more computationally feasible it is to integrate data
coherently (preserving phase information) over long observation times, for reasons
explained below.

In theory, a definitive continuous-wave source detection can be
accomplished with a single gravitational wave detector because the source
remains on, allowing follow-up verification of the signal strength and of
the distinctive Doppler 
modulations of signal frequency due to the Earth's motion (discussed below). 
Hence a relatively large number of CW searches were carried out with both bar
detectors and interferometer prototypes in the decades even before the major
1st-generation interferometers began collecting data, with many additional
searches carried out in Initial LIGO and Virgo data~\cite{bib:rilesppnp}.
In the following, we focus mainly on the most recent results 
from Advanced LIGO's first observing run O1. 

\subsection{Targeted and narrowband searches for known pulsars}
\label{sec:targeted}

In {\it targeted} searches for known pulsars using measured ephemerides from radio, optical, X-ray or $\gamma$-ray
observations valid over the gravitational wave observation time, one can apply
corrections for phase modulation (or, equivalently, frequency modulation) due
to the motion of the Earth (daily rotation and orbital motion), and in the case of binary
pulsars, for additional source orbital motion. For the Earth's motion, one has
a daily relative frequency modulation of $v_{\rm rot}/c\approx10^{-6}$ and a much
larger annual relative frequency modulation of $v_{\rm orb}/c\approx10^{-4}$. 
The pulsar astronomy community has developed a powerful and mature software infrastructure for
measuring ephemerides and applying them in measurements, using the TEMPO~2 program~\cite{bib:tempo}.
The same physical corrections for the Sun's, Earth's and Moon's motions (and for the motion of other planets),
along with general relativistic effects
including gravitational redshift in the Sun's potential and Shapiro delay for
waves passing near the Sun, have been incorporated into the LIGO and Virgo software
libraries~\cite{bib:lal}. 

Consider the model of a rotating rigid triaxial ellipsoid (model for a neutron star), for which
the strain waveform detected by an interferometer can be written as
\begin{equation}
\label{eqn:cwhdefinition}
h(t) \quad=\quad F_+(t,\psi)\,h_0{1+\cos^2(\iota)\over2}\,\cos(\Phi(t)) 
\>+\> F_\times(t,\psi)\,h_0\,\cos(\iota)\,\sin(\Phi(t)),
\end{equation}
where $\iota$ is the angle between the star's spin direction and the propagation
direction $\hat k$ of the waves (pointing toward the Earth), 
where $F_+$ and $F_\times$ are the (real) detector antenna pattern response factors
($-1 \le F_+,F_\times \le 1)$ to the $+$ and $\times$ polarizations. $F_+$ and $F_\times$ 
depend on the orientation of the detector and the source, and on 
the polarization angle $\psi$~\cite{bib:cwtargeteds1}. Here, $\Phi(t)$ is
the phase of the signal.

The phase evolution of the signal can be usefully expanded
as a truncated Taylor series:
\begin{equation}
\label{eqn:phasedefinition}
\Phi(t) \quad = \quad \phi_0 + 2\,\pi\left[f_s(T-T_0) + {1\over2}\dot f_s(T-T_0)^2 + 
{1\over6}\ddot f_s(T-T_0)^3\right],
\end{equation}
where
\begin{equation}
\label{eqn:phaseevolution}
T \quad = \quad t + \delta t \quad = \quad t - {\vec r_d\cdot\hat k\over c}+ \Delta_{E\odot}-\Delta_{S\odot}.
\end{equation}
Here, $T$ is the time of arrival of a signal at the solar system barycenter (SSB),
$\phi_0$ is the phase of the signal at fiducial time $T_0$, $\vec r_d$ is the position
of the detector with respect to the SSB, and $\Delta_{E\odot}$ and $\Delta_{S\odot}$ 
are solar system Einstein and Shapiro time delays, respectively~\cite{bib:taylorssb}.
In this expression $f_s$ is the nominal instantaneous frequency of the gravitational
wave signal [twice the star's rotation frequency~\cite{bib:zimmermannszedenits} for a signal created by a rotating star's
non-zero ellipticity, as in equations~(\ref{eqn:ellipticity}-\ref{eqn:hexpected})].

Various approaches have been used in targeted searches in LIGO and Virgo data to date:
1) A time-domain heterodyne method~\cite{bib:dupuiswoan} in which Bayesian posteriors are determined on
the signal parameters that govern absolute phase, amplitude and
amplitude modulations; 2) a matched-filter method in which marginalization
is carried out over unknown orientation parameters (``\fstatistic'')~\cite{bib:jks,bib:gstatisticmethod}; 
and 3) a Fourier-domain determination of a ``carrier'' strength along with the strengths
of two pairs of sidebands created by amplitude modulation from the Earth's sidereal
rotation of each detector's antenna pattern (``5-Vector'' method)~\cite{bib:5vector,bib:5vectorupdates}.

\begin{figure}[t!]
\begin{center}
%\vspace{-0.7in}
\includegraphics[width=13.cm]{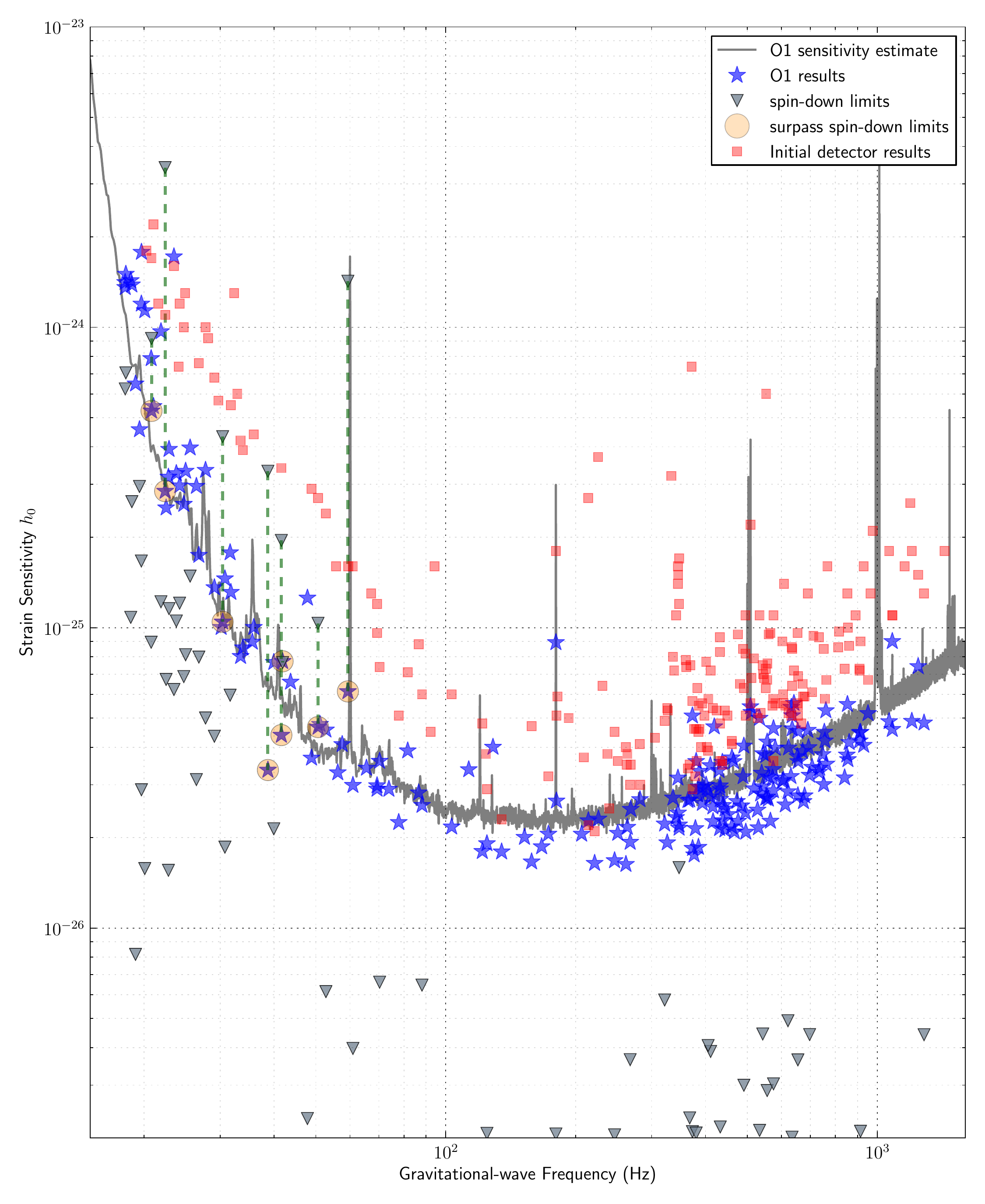}
%\vspace{-0.85in}
\caption{Upper limits on $h_0$ for known pulsars from searches in the 
LIGO O1 data~\cite{bib:cwtargetedO1}. The gray band shows the {\it a priori} estimated 
sensitivity range of the search. Also plotted 
are the lowest upper limits from searches in initial LIGO and Virgo data.}
\label{fig:cwtargetedO1}
\end{center}
\end{figure}

Results from searches of each type~\cite{bib:cwtargetedO1} are shown in Figure~\ref{fig:cwtargetedO1}, where
method 1) was applied to 200 stars, and methods 2) and 3) were applied to 11 and 10 stars,
respectively, for which the spin-down limit (equation~\ref{eqn:spindownlimit}) was likely
to be beaten or approached, given the detector sensitivity. Highlights of these searches include
setting a lowest upper limit on strain amplitude of {$\bf1.6\times10^{-26}$}
(J1918-0642),
setting a lowest upper limit on ellipticity of {$\bf1.3\times10^{-8}$} (J0636+5129) and beating
the spin-down limit on 8 stars (J0205+6449, J0534+2200, J0835-4510, J1302-6350, J1813-1246, J1952+3252,
J2043+2740, J2229+6114).
Perhaps the most notable result was setting
an upper limit on the Crab pulsar's (J0534+2200) energy loss to gravitational radiation at a level
of {\bf0.2\%} of the star's total rotational enegy loss inferred from measured rotational spin-down.

These upper limits assume the correctness of General Relativity in that antenna pattern calculations
used in the searches assume two tensor polarizations in strain. Alternative theories of gravity can,
in principle, support four additional polarizations (two scalar and two vector modes), which would
lead to different antenna pattern sensitivities~\cite{bib:TGRmethod}. Searches have been carried out for evidence of signals
from the 200 targeted pulsars above exhibiting these other polarizations, using the heterodyned data products.
In no case was significant evidence of a non-standard signal seen, and upper limits were placed~\cite{bib:TGRO1}.

The targeted-search upper limits in Figure~\ref{fig:cwtargetedO1} assume a fixed phase relation between stellar rotation (measured by
electromagnetic pulses) and gravitational wave emission ($f_s=f_{\rm rot}$). To allow for a
more general scenario, such as slight differential rotation of EM- and GW-emitting regions,
searches have also been carried out for signals very near in parameter space to those expected
from an ideal phase relation. These so-called ``narrowband'' searches allow a relative frequency deviation
of O($10^{-3}$). Results from searches for 11 stars with expected sensitivities near
the spin-down limits have been obtained from O1 data~\cite{bib:narrowbandO1}. In general, these limits are expected
and found to be higher than the corresponding upper limits from targeted searches above because the
increased parameter space search implies an additonal trials factor. Nonetheless, this narrowband
search beat the spin-down limit on the Crab (J0534+2200), Vela (J0835-4510) and J2229+6114.

\subsection{Directed searches for isolated stars}
\label{sec:directedisolated}

Unlike targeted searches, where the phase evolution of the signal
is (assumed to be) known precisely enough to permit a coherent integration over
the full observation time, in a {\it directed} search one has limited or no
information about the phase evolution of the source, while knowing precisely
the sky location of the star. 
The implied parameter space volume of the search will
then depend sensitively upon the assumed age of the star. For a very young
pulsar, one must search over not only the frequency and first frequency derivative
(spin-down), but also over the second and possibly higher derivatives.

To understand the scaling, imagine carrying out a coherent search, where
one wishes to maintain phase coherence over the observation span $T_{\rm obs}$ of no
worse than some error $\Delta\Phi$. From equation~(\ref{eqn:phasedefinition}), one needs
to search over $f_s$ in steps proportional to $1/T_{\rm obs}$,
over $\dot f_s$ in steps proportional to $1/T_{\rm obs}^2$, and
over $\ddot f_s$ in steps proportional to $1/T_{\rm obs}^3$. Hence a search
that requires stepping in $\ddot f_s$ will have a parameter space
volume proportional to $T_{\rm obs}^6$, with search time through the data incurring
another power of $T_{\rm obs}$ in total computing cost (although accounting for the
discreteness of the search over $\ddot f_s$ reduces the power scaling for moderate $T_{\rm obs}$ values).
Hence, even when the source direction is precisely
known, the computational cost of a {\it coherent} search over $f_s$, $\dot f_s$ and $\ddot f_s$
grows extremely rapidly with observation time. One can quickly exhaust all available computing capacity
by choosing to search using a $T_{\rm obs}$ value that coincides with a full data
set, \eg, two years. In that case, one may simply choose the largest $T_{\rm obs}$ value
with an acceptable computing cost, or one may choose instead a {\it semi-coherent}
strategy of summing strain powers from many smaller time intervals, as discussed below
in the context of all-sky searches. Both approaches have been used in recent years in
analysis of Initial and Advanced LIGO data. As of this article's submission, no
directed searches of Advanced LIGO data have been published, so here the focus will
be on the final searches carried out in Initial LIGO data with a brief discussion of future
prospects.

\begin{figure}[t!]
\begin{center}
\includegraphics[width=12.cm]{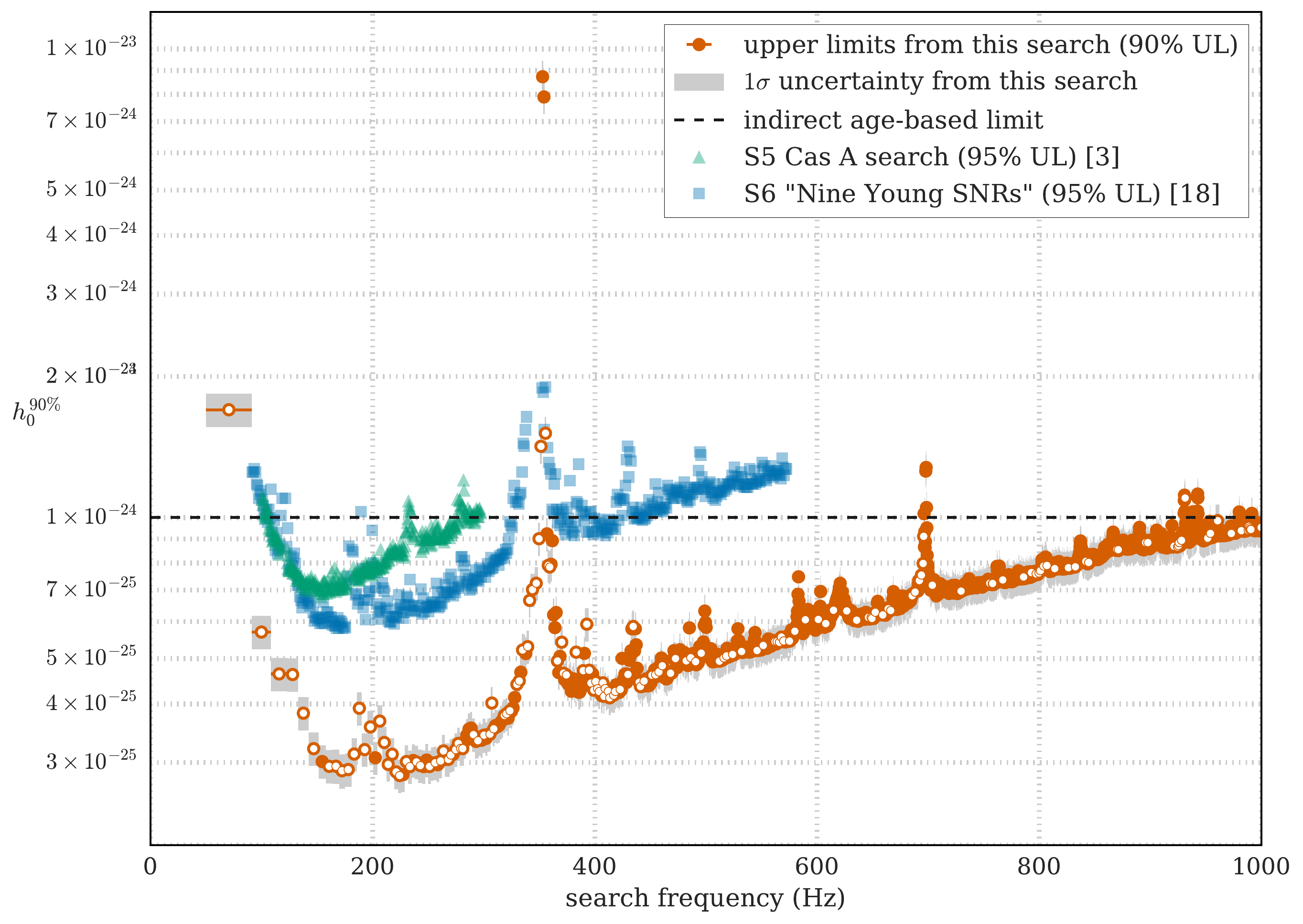}
\caption{Upper limits on $h_0$ for Cassiopeia A from Initial LIGO S6 data,~\cite{bib:directedE@HS6} shown with
  previous limits using the same or older data. Open circles indicate bands with partial contamination from instrumental disturbances.}
\label{fig:S6CasAE@H}
\end{center}
\end{figure}

The coherent approach over tractable intervals~\cite{bib:cwcasamethod} was applied to searches in the data from
the last Initial LIGO data run (S6) for nine young
Supernova remnants~\cite{bib:S6NineSNRs} and to a possible source at the core of the
globular cluster NGC 6544~\cite{bib:S6NGC6544}.
\newif\ifshowsnruls\showsnrulsfalse
\ifshowsnruls
Figure~\ref{fig:S6SNRULs} shows 95\%\ confidence
upper limits obtained from the SNR search. Comparable upper limits are obtained from S6 data for NGC 6544.
\fi
In addition, the semi-coherent approach was applied in a computationally intensive Einstein@Home (see section~\ref{sec:allsky})
S6 search for continuous gravitational waves from the X-ray
source at the center of the Cassiopeia supernova remnant,~\cite{bib:directedE@HS6} which is especially interesting because of its youth,
as discussed in section~\ref{sec:neutronstarsources}. Figure~\ref{fig:S6CasAE@H} shows the results
of the Cas~A search in comparison with the results from earlier searches, including from the coherent S6 search
above,~\cite{bib:S6NineSNRs} from which it can be seen that the semi-coherent search,
which exploits the full data set, achieves nearly a factor
of two improvement in strain sensitivity. Note that these limits are roughly an order of magnitude
higher, on the whole, than the limits achieved in the targeted search (Figure~\ref{fig:cwtargetedO1})
using known ephemerides. 
 
%\fi
 
Another approach~\cite{bib:xcorrmethod1} for directed searches is based on 
cross correlation of independent data streams. The most straightforward method
defines bins in detector-frame frequency and uses short coherence times, as 
in directional searches for stochastic gravitational radiation,~\cite{bib:radiometermethod,bib:DirectedStochasticO1}
which can be used to search for both isolated and binary sources, albeit with limited sensitivity.
One can use finer frequency binning, however, when correcting explicitly for Doppler modulation of the signal.
Cross-correlation methods are especially robust against wrong assumptions about phase
evolution and are attractive in searching for a very young object, such as
a hypothetical neutron star remaining from Supernova 1987A (see Ref.~\refcite{bib:ashtonprixjones} for
a discussion of potential degradation of coherent searches from neutron star glitches).
In fact, a cross-correlation search for SN 1987A, including demodulation for effects from the motion
of the Earth,~\cite{bib:xcorrmethod2}
was carried out in Initial LIGO data~\cite{bib:xcorrS5} Recent application of cross-correlation methods to
directed searches for binary sources will be discussed in the next section (\ref{sec:directedbinary}).

Directed searches for particular sources require making choices, that is, to prioritize among a
wide set of potential targets in deciding how best to apply computational resources and analyst time.
Recent work~\cite{bib:targetchoice1,bib:targetchoice2} has taken a Bayesian approach to address this
problem, one that may be generalized to parameter choices in both directed and all-sky searches.

\subsection{Directed searches for binary stars}
\label{sec:directedbinary}

For known binary pulsars with measured timing ephemerides,
targeted searches work well, and upper limits have been reported for many 
stars, as described in section~\ref{sec:targeted}.
But searching for known (possibly accreting) binary neutron stars
not exhibiting pulsations  or for entirely unknown binary stars
once again significantly increases the parameter space,
relative to the corresponding isolated star searches,
posing new algorithmic challenges and computing costs. 

Because of its high X-ray flux and the torque-balance
relation for low-mass X-ray binaries [equation~(\ref{eqn:torquebalance})],
Scorpius X-1 is thought to be an especially promising search
target for advanced detectors and has been the subject
of multiple searches in Initial and Advanced LIGO data. From equation~(\ref{eqn:torquebalance}),
one expects a strain amplitude limited by~\cite{bib:cwfstats2,bib:ScoX1MDC1}
\begin{equation}
  h \quad \sim \quad (3.5\times10^{-26})\,\left({600\>{\rm Hz}\over f_{GW}}\right)^{1\over2}.
  \label{eqn:torquebalance2}
\end{equation}
While Sco X-1's rotation frequency
remains unknown, its orbital period is well measured,~\cite{bib:scox1period}
which allows substantial reduction in search space. 

\begin{figure}[t!]
\begin{center}
\includegraphics[width=12.cm]{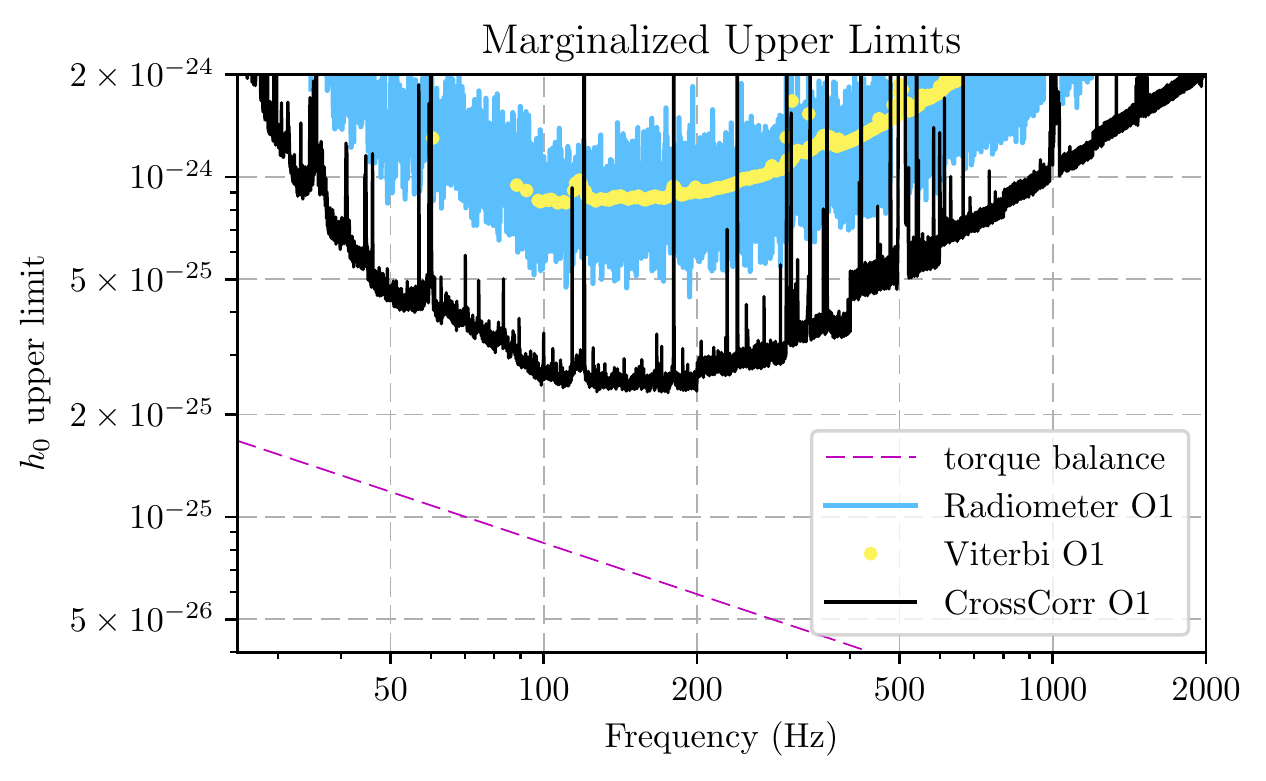}
\caption{Upper limits on $h_0$ for Scorpius X-1 from Advanced LIGO O1 data, using several different search methods.~\cite{bib:CrossCorrO1} The dashed line indicates the torque-balance benchmark.}
\label{fig:O1CrossCorrScoX1}
\end{center}
\end{figure}

Searches for Sco X-1 in O1 data have been carried out with several
methods: 1) a ``Sideband'' method~\cite{bib:sidebandmethod1,bib:sidebandmethod2,bib:sidebandviterbi,bib:sidebandO1}
based on summing power in orbital sideband frequencies;
2) a non-demodulated cross-correlation methods~\cite{bib:DirectedStochasticO1}
and 3) a demodulated cross-correlation method~\cite{bib:xcorrmethod3,bib:CrossCorrO1}.
The demodulated cross-correlation method has proven to be the most sensitive method to date in such searches,
as expected from a previous mock data challenge~\cite{bib:ScoX1MDC1} including these methods
and others,~\cite{bib:twospectmethod,bib:twospectsumming,bib:twospectdirectedmethod,bib:twospectS6,bib:polynomial}
and as shown in Figure~\ref{fig:O1CrossCorrScoX1}, although computationally intensive methods using the
\fstatistic\ may eventually improve upon it~\cite{bib:StackedFstatScoX1Method}. The strain upper limits
shown in Figure~\ref{fig:O1CrossCorrScoX1} do not reach as low as the torque-balance benchmark in
equation~(\ref{eqn:torquebalance2}), but at Advanced LIGO design sensitivity and with longer data runs,
future searches should begin to probe at least the low-frequency range of this benchmark.
One complication in Sco X-1 searches is potential spin wandering due to fluctuations in accretion
from its companion,~\cite{bib:spinwandering} which limits the length of a coherence time that can
be assumed safe for a signal template.
One previous fully coherent search~\cite{bib:sidebandS5} restricted its coherence length to 10 days,
to be conservative. Semi-coherent and cross-correlation
methods~\cite{bib:sidebandviterbi,bib:twospectdirectedmethod,bib:radiometermethod,bib:xcorrmethod3}
should be more robust against wandering, an expectation to be tested soon in a new mock data challenge.

It should be stated that obtaining more definitive information on
the rotation frequency of Sco X-1 could potentially make the
difference between missing and detecting its gravitational
waves in advanced detector data, by reducing the statistical 
trials factor and thereby the threshold needed to identify
an interesting outlier. {\it More intensive measurements and
analysis of Sco X-1 X-ray emission could yield a dramatic scientific payoff.}

\subsection{All-sky searches for isolated stars}
\label{sec:allsky}

In carrying out {\it all-sky} searches for unknown neutron stars, 
the computational considerations grow worse. The corrections for
Doppler modulations and antenna pattern modulation due to the Earth's 
motion must be included, as for the targeted and directed searches,
but the corrections are sky-dependent, and the spacing of the
demodulation templates is dependent upon the inverse of
the coherence time of the search. Specifically, for a coherence time $T_{\rm coh}$
the required angular resolution is~\cite{bib:cwallskys4}
\begin{equation}
\label{eqn:angres}
\delta\theta \quad \approx \quad {0.5\, {\rm c}\, \delta f\over f\,[v\sin(\theta)]_{\rm max}},
\end{equation}
where $\theta$ is the angle between the detector's velocity relative
to a nominal source direction, where the maximum relative frequency shift 
$[v\sin(\theta)]_{\rm max}/c\approx10^{-4}$, and where $\delta f$
is the size of the frequency bins in the search. For $\delta f=1/T_{\rm coh}$,
one obtains:
\begin{equation}
\delta\theta \quad \approx \quad 9\times 10^{-3}\>{\rm rad}\>\left({30\>{\rm minutes}\over T_{\rm coh}}\right)
\left({300\>{\rm Hz}\over f_s}\right),
\end{equation}
where $T_{\rm coh}$ = 30 minutes has been used in several all-sky searches to date.
Because the number of required distinct points on the sky scales like $1/(\delta\theta)^2$,
the number of search templates scales like $(T_{\rm coh})^2(f_s)^2$ for a fixed signal frequency $f_s$.
Now consider attempting a search with a coherence time of 1 year for a
signal frequency $f_s=1$ kHz. One obtains $\delta\theta\sim0.3$ $\mu$rad and
a total number of sky points to search of O(10$^{14}$) -- again, for a fixed
frequency. Adding in the degrees of freedom to search over ranges in 
$f_s$, $\dot f_s$ and $\ddot f_s$ makes a fully coherent 1-year all-sky
search utterly impractical, given the Earth's present total computing capacity.

As a result, tradeoffs in sensitivity must be made to achieve tractability
in all-sky searches. The simplest tradeoff is to reduce the observation
time to an acceptable coherence time.
It can be more attractive, however, to reduce the coherence time still further
to the point where the total observation time is divided into $N=T_{\rm obs}/T_{\rm coh}$,
segments, each of which is analyzed coherently and the results added incoherently
to form a detection statistic. One sacrifices intrinsic sensitivity per 
segment in the hope of compensating (partially) with the 
increased statistics from being able to use more total data, as discussed above for
directed searches. In practice, for realistic data observation spans (weeks or longer), the semi-coherent
approach gives better sensitivity for fixed computational cost and hence has been used
extensively in all-sky searches. One finds a
strain sensitivity (threshold for detection) that scales approximately as the inverse fourth root 
of $N$~\cite{bib:cwallskys2}. Hence, for a fixed observation time, the sensitivity degrades
roughly as $N^{1\over4}$ as $T_{\rm coh}$ decreases. This degradation is a price one pays
for not preserving phase coherence over the full observation 
time, in order to make the search computationally tractable. An important virtue of
semi-coherent searches methods is robustness with respect to deviations of a signal from an assumed coherent model.

Various semi-coherent algorithmic approaches have been tried, most based in some way on the
``Stack Slide'' algorithm~\cite{bib:stackslide1,bib:stackslide2,bib:stackslide3} in which the power from Fourier
transforms over each coherently analyzed segment is stacked on each other after
sliding each transform some number of bins to account for Doppler modulation of
the source frequency.
One algorithm is a direct implementation of this idea called StackSlide~\cite{bib:stackslideimplementation}.
Other implementations~\cite{bib:houghmethod,bib:freqhough1} are based on the Hough transform
approach,~\cite{bib:houghibm1,bib:houghibm2} 
in which for each segment a detection statistic is compared to a threshold and given
a weight of 0 or 1 (later refined to include adaptive non-unity weights, to account for
variations in noise and detector antenna pattern~\cite{bib:adaptivefreqhough,bib:adaptiveskyhough}).
The sums of those weights are accumulated in parameter space ``maps,''
with high counts warranting follow-up. The Hough approach offers, in principle, somewhat
greater computational efficiency from reducing floating point operations, but its greater
value lies in its robustness against non-Gaussian artifacts~\cite{bib:cwallskys4}.
The Hough approach has been implemented in two distinct search pipelines, the ``Sky Hough''~\cite{bib:houghmethod,bib:skyhough1} and
``Frequency Hough''~\cite{bib:freqhough1,bib:freqhough2,bib:freqhough3} programs, named after the different parameter spaces chosen in which to
accumulate weight sums.
Another implementation, known as PowerFlux,~\cite{bib:powerflux1,bib:powerflux2,bib:loosecoherence,bib:universalstatistic}
improves upon the StackSlide method by weighting segments by the inverse variance
of the estimated (usually non-stationary) noise and by searching explicitly over
different assumed polarizations while including the antenna pattern correction factors
in the noise weighting. Yet another method uses coincidences among \fstatistic\ outliers in
multiple time segments typically longer than those used in the semi-coherent approaches.\cite{bib:coincfstat1,bib:coincfstat2}

The deepest searches achieved to date have stacked \fstatistic\ values over time segments
semi-coherently and have required the computational resources of the distributed computing
project Einstein@Home~\cite{bib:cwe@hs4} which uses the same software infrastructure (BOINC)~\cite{bib:boinc}
developed for the Seti@Home project~\cite{bib:seti@home}. The algorithms used in Einstein@Home have
evolved steadily in sophistication and sensitivity over the last decade.
Particular improvements have included search setup optimization~\cite{bib:stackslide3,bib:shaltev},
more efficient semi-coherent stacking and template placement,~\cite{bib:prixtemplate,bib:pletsch,bib:pletschallen,bib:wetteprix,bib:wettetemplate,bib:wettee@h1,bib:wettee@h2}
automated vetoing of instrumental lines,~\cite{bib:lineveto1,bib:lineveto2} and 
hierarchical outlier followup and veto.~\cite{bib:prixshaltev,bib:shaltevetal,bib:papafollowup,bib:singh,bib:zhu}
  
\begin{figure}[htb]
\begin{center}
  \includegraphics[width=5in]{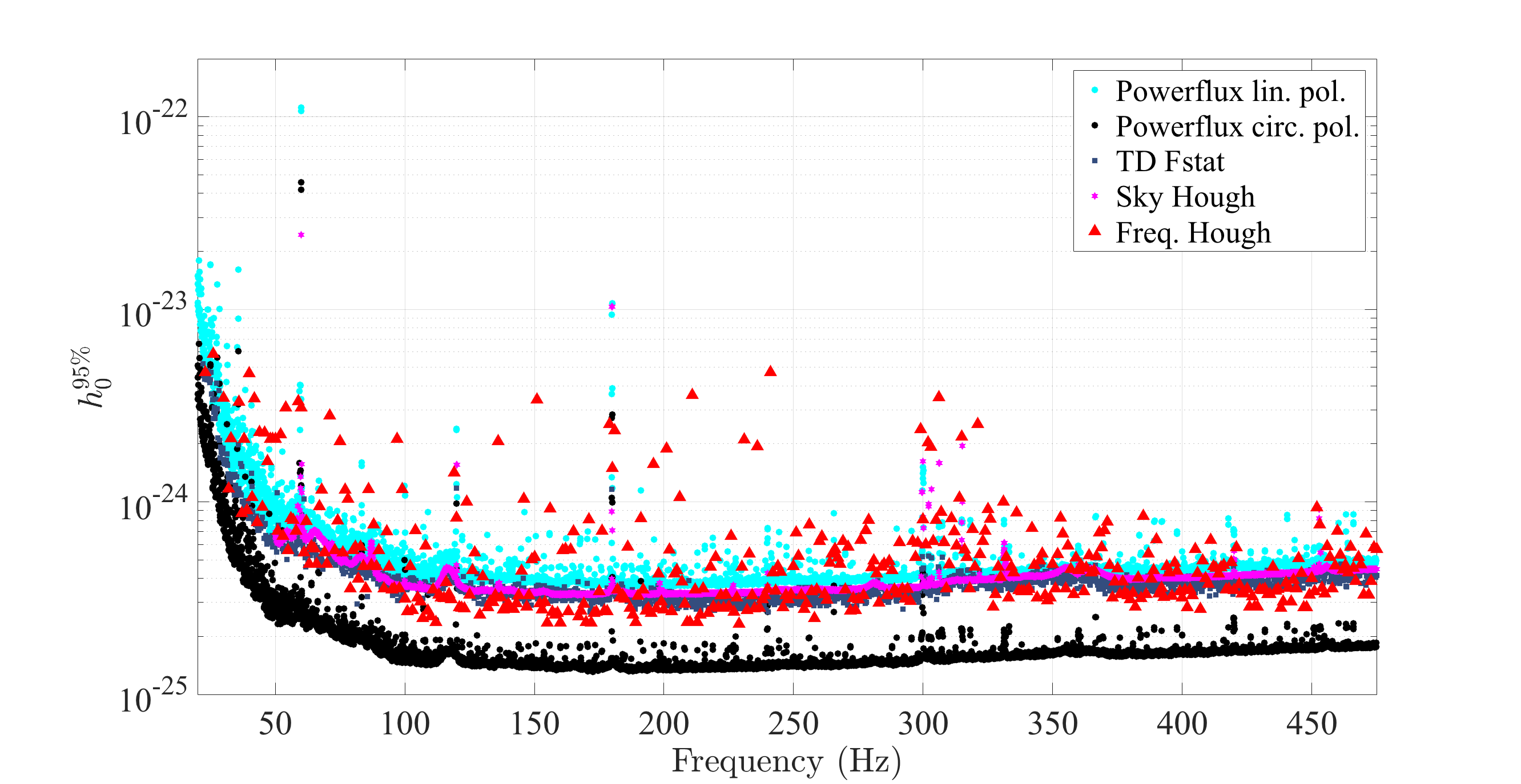}
\vspace{-\abovedisplayskip}
\caption{All-sky upper limits on unknown sources of continuous waves from the Advanced LIGO O1 search,~\cite{bib:O1allskylowfreq}
  based on five different search pipelines, as described in the text. Four pipelines cover the band 20-475 Hz, and the
  deepest search (Einstein@Home) covers the low-frequency band 20-100 Hz. The upper limits shown from the PowerFlux pipeline
  indicate the extremes between circular polarization (most optimistic) and worst-case linear polarization.}
\label{fig:O1AllskyLowfreq}
\end{center}
\end{figure}

\begin{figure}[htb]
\begin{center}
  \includegraphics[width=5in]{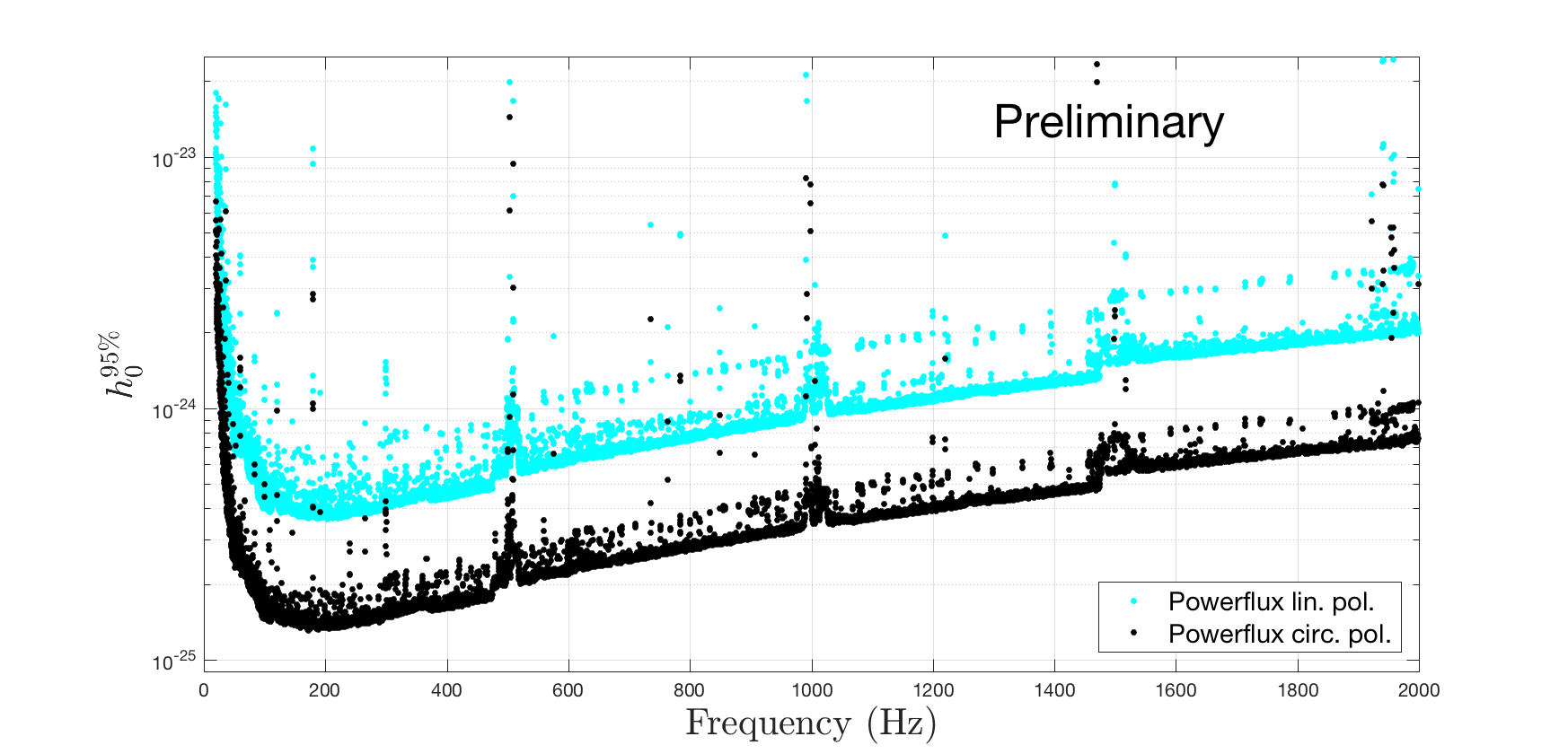}
\vspace{-\abovedisplayskip}
 \caption{Preliminary all-sky upper limits from the PowerFlux search pipeline over the band 20-2000 Hz on unknown sources of continuous waves from the Advanced LIGO O1 search~\cite{bib:VladimirAPS2017}. As in the preceding figure, upper limits for circular and linear polarizations are shown separately.}
\label{fig:O1PowerFluxAllfreq}
\end{center}
\end{figure}

A recent comparison of these methods was carried out via a mock
data challenge using initial LIGO data,~\cite{bib:allskymdc} and these methods have been applied to published
searches of the Advanced LIGO O1 data set~\cite{bib:O1allskylowfreq,bib:O1E@H}. Figure~\ref{fig:O1AllskyLowfreq} shows upper limits 
on strain from five searches (four covering 20-475 Hz and the deeper Einstein@Home covering 20-100 Hz with a smaller spin-down range).
{\it Preliminary} upper limits,~\cite{bib:VladimirAPS2017} from one of the pipelines (PowerFlux) are also shown for a broader frequency range in Figure~\ref{fig:O1PowerFluxAllfreq},
with results of comparable sensitivity from multiple pipelines expected to be published soon.

\subsection{All-sky searches for binary stars}
\label{sec:allskybinary}

Given the computational difficulty in carrying out a search
over the two unknown orbital parameters of a known binary star
with known period and assumed circular orbit, it should come as no surprise that carrying
out a search over three or more unknown orbital parameters for an
unknown binary star anywhere in the sky is especially challenging.
Several methods have been proposed for carrying out such an all-sky binary search,
which approach the problem from opposite extremes. The first method, which has been
used in a published search of Initial LIGO S6 data~\cite{bib:twospectresultsS6} is
known as TwoSpect~\cite{bib:twospectmethod}. The program carries out a
semi-coherent search over an observation time long compared to
the maximum orbital period considered, while using coherence times short with
respect to the orbital period. Fourier transforms are carried
out over each row (fixed frequency bin) in a $\sim$year-long
spectrogram and the resulting frequency-frequency plot searched
for characteristic harmonic patterns. Another developed pipeline, known
as Polynomial,~\cite{bib:polynomial} searches coherently using
matched filters over an observation time short compared to the minimum orbital period
considered. A bank of frequency polynomials in time is used for
creating the matched filters, where for a small segment of an orbit,
the frequency should vary as a low-order polynomial. Other proposed methods, which
offer potentially substantial computational savings at a cost in sensitivity, include 
autocorrelations in the time-frequency plane~\cite{bib:vicereautocorr} and
stochastic-background techniques using skymaps with sidereal-day folding~\cite{bib:stochfolding}.
Results from one or more of these search methods applied to Advanced LIGO O1 data are expected in
early 2018. 

\section{Prospects for the Future}
\label{sec:future}

Over the next several years, the Advanced LIGO and Virgo detectors are expected
to approach their design sensitivities in strain, improving by about another factor
of three over current broadband sensitivities~\cite{bib:obsscenario}. This
dramatic improvement in strain increases the volume of the galaxy searched
with CW pipelines by a factor between 9 and 27, depending on assumed neutron
star ellipticies, signal frequencies, \etc, thereby increasing detection likelihood.
As search ranges approach the dense galactic core, detection chances may rise more
rapidly. In parallel to detector improvement, algorithms continue to improve, as researchers
find more effective tradeoffs between computational cost and detection efficiency,
while Moore's Law, including Graphical Processing Unit (GPU) exploitation, ensures
increased computing resources for searches. All of these trends are encouraging
for successful CW detection.

At the same time, theoretical uncertainties on what sensitivity is needed for the
first CW detection are very large. While the spin-down limits based on gravitar assumptions
and on either energy conservation or known age have been beaten for a handful
of sources and will be beaten for more sources in the coming years, the gravitar
model is surely optimistic -- most stellar spin-downs are likely dominated by
electromagnetic interactions. Whether the first detection is imminent or still many
years distant remains unclear.

Electromagnetic astronomers could prove pivotal in hastening detection by identifying
new nearby or young neutron stars, or discovering pulsations from known stars, perhaps most usefully
from the accreting Sco X-1 system. Given the computational challenges of most CW searches,
narrowing the parameter space of a search exploiting electromagnetic observations could make
the difference between a gravitational wave miss and a discovery.

\section*{Acknowledgments}

The author is deeply grateful to colleagues in the LIGO Scientific Collaboration
and Virgo collaboration Continuous Waves Search Group for close collaboration
from which he has benefited in preparing this article. The author also thanks 
Cristiano Palomba, David Keitel, Maria Alessandra Papa, Karl Wette and John Whelan
for helpful suggestions concerning the manuscript
and thanks LIGO, Virgo and the Albert Einstein Institute
for the use of figures. This work was supported in part
by National Science Foundation Award PHY-1505932.

%\section*{References}

\end{document}